\documentclass[aps,prd,reprint,nobibnotes,twocolumn,superscriptaddress,bibliography]{revtex4-2}
\usepackage{amsmath}
\usepackage{amssymb}
\usepackage{bm}
\usepackage{graphicx}
\usepackage{color}
\usepackage{dcolumn}
\usepackage{amsfonts}
\usepackage{mathrsfs}
\usepackage{xspace}
\usepackage{epstopdf}
\usepackage{longtable}
\usepackage{multirow}
\usepackage{float}
\usepackage{comment}
\usepackage[colorlinks=true, letterpaper=true, pdfstartview=FitV, linkcolor=blue, citecolor=blue, urlcolor=blue]{hyperref}

\providecommand{\tabularnewline}{\\}

\begin{document}
\title{Chiral phonons in a square lattice}

\author{Qianqian Wang}
\affiliation{Research Laboratory for Quantum Materials, Singapore University of
	Technology and Design, Singapore 487372, Singapore}

\author{Si Li}
\affiliation{School of Physics, Northwest University, Xi'an 710069, China}

\author{Jiaojiao Zhu}
\affiliation{Research Laboratory for Quantum Materials, Singapore University of
	Technology and Design, Singapore 487372, Singapore}

\author{Hao Chen}
\affiliation{NNU-SULI Thermal Energy Research Center (NSTER) \& Center for Quantum Transport and Thermal Energy Science (CQTES), School of Physics and Technology, Nanjing Normal University, Nanjing 210023, China}
\affiliation{Division of Physics and Applied Physics, School of Physical and Mathematical Sciences, Nanyang Technological University, Singapore 637371, Singapore}

\author{Weikang Wu}\email{weikang.wu@ntu.edu.sg}
\address{Division of Physics and Applied Physics, School of Physical and Mathematical Sciences, Nanyang Technological University, Singapore 637371, Singapore}
\address{Research Laboratory for Quantum Materials, Singapore University of Technology and Design, Singapore 487372, Singapore}

\author{Weibo Gao}\email{wbgao@ntu.edu.sg}
\address{Division of Physics and Applied Physics, School of Physical and Mathematical Sciences, Nanyang Technological University, Singapore 637371, Singapore}

\author{Lifa Zhang}
\email{phyzlf@njnu.edu.cn}
\affiliation{NNU-SULI Thermal Energy Research Center (NSTER) \& Center for Quantum Transport and Thermal Energy Science (CQTES), School of Physics and Technology, Nanjing Normal University, Nanjing 210023, China}

\author{Shengyuan A. Yang}
\affiliation{Research Laboratory for Quantum Materials, Singapore University of
	Technology and Design, Singapore 487372, Singapore}

\begin{abstract}
	Chiral phonons were initially proposed and further verified experimentally in two-dimensional (2D) hexagonal crystal lattices. Many intriguing features brought about by chiral phonons are attributed to the pseudo-angular momenta which are associated with the threefold rotational symmetry of hexagonal lattices.
	Here, we go beyond the hexagonal crystals and investigate the chiral phonons in systems with fourfold rotational symmetry. We clarify the symmetry requirements for the emergence of chiral phonons in both 2D square lattices and 3D tetragonal lattices. For 2D, the realization of $C_4$ chiral phonons requires the breaking of time-reversal symmetry; while for 3D, they can exist on the $C_4$-invariant path in a chiral tetragonal lattice. These phonons have the advantage that they can be more readily coupled with optical transitions, which facilitates their experimental detection. We demonstrate our idea via model analysis and first-principles calculations of concrete materials, including the MnAs monolayer and the $\alpha$-cristobalite. Our work reveals chiral phonons beyond the hexagonal lattices and paves the way for further exploration of chiral phonon physics in square/tetragonal materials and metamaterials.
    \end{abstract}
\maketitle	

\section{Introduction}

As collective vibrations of crystal lattices, phonons play an important role in solid-state physics, underpinning many fundamental phenomena ranging from specific heat to superconductivity. In 2015, the concept of chiral phonons was proposed, for which the vibration modes acquire a definite sense of chirality, either right-handed or left-handed~\cite{zhang2015chiral}. Chiral phonons naturally exhibit selective coupling with other chiral quantities/excitations, such as circularly-polarized light~\cite{zhang2015chiral,zhu2018observation,chen2019entanglement,li2019momentum,chen2021propagating}, magnetization~\cite{hamada2018phonon}, phonon Berry curvature~\cite{zhang2010topological,liu2018berry}, and chiral structures~\cite{chen2021chiral}, which are expected to generate novel physical effects.  As a result, chiral phonons have been attracted great interest in recent years. Combined theoretical and experimental efforts have revealed their important contributions to various optical and excitonic effects in two-dimensional (2D) semiconductors~\cite{zhu2018observation,chen2018chiral,li2020phonon},
magneto-optical response in topological semimetals~\cite{yin2021chiral}, electronic or structural phase transitions~\cite{mankowsky2016non,forst2015mode}, valley phonon Hall effect~\cite{he2020valley}, and etc~\cite{jotzu2014experimental,grissonnanche2020chiral,delhomme2020flipping,romao2019anomalous,pandey2018symmetry,ren2021phonon,long2020unsupervised}.

In studying chiral phonons, the focus is on the phonon modes at high-symmetry points of the Brillouin zone (BZ). This is because besides sizable intrinsic angular momenta, due to symmetry, these modes also possess well defined pseudo-angular momenta (PAM)~\cite{zhang2015chiral}, which can enable their selective coupling to optical transitions and other excitations. Indeed, the first experimental confirmation of chiral phonons in 2018 was based on the infrared circular dichroism in monolayer WSe$_2$~\cite{zhu2018observation}, where the chiral phonons at $K$ and $K'$ high-symmetry points of the hexagonal BZ selectively participate in the intervalley optical transitions.

It follows that the properties of chiral phonons should depend on the type of the crystal lattice, since different lattices have different high-symmetry points with different constraints on the phonon modes. Previous studies on chiral phonons have covered a variety of lattice models, such as the honeycomb lattice~\cite{zhang2015chiral}, Kekule lattice~\cite{liu2017pseudospins}, kagome lattice~\cite{chen2019chiral}, $\sqrt{3}\times\sqrt{3}$ honeycomb superlattice~\cite{xu2018nondegenerate}, and also realistic materials and heterostructures, such as graphene/BN~\cite{gao2018nondegenerate}, 2D transition metal dichalcogendies~\cite{zhu2018observation}, {WN$_{2}$}~\cite{chen2021propagating}, and $\alpha$-quartz~\cite{chen2021chiral}. However, one notes that all those works are limited to the hexagonal crystal system, there the PAM of chiral phonons are associated with the threefold rotational symmetry $C_3$.

Then, a question naturally arises: \emph{Can chiral phonons be extended beyond the hexagonal crystal system?}

In this work, we address the above question. Since the translational symmetry of a lattice is compatible with only 2-, 3-, 4-, and 6-fold rotational axes, it appears that a natural candidate is a system with fourfold rotational axis. In 2D, this corresponds to the square lattice.  We clarify the underlying symmetry condition for the emergence of chiral phonons in a square lattice with $C_4$ symmetry. Particularly, since the $C_4$-invariant points in the BZ are all time-reversal-invariant momentum (TRIM) points, we show that chiral phonons there can only appear when the time reversal symmetry $\mathcal{T}$ is broken.
We demonstrate the idea explicitly by a model calculation. We show that different from previous cases, here, chiral phonons can appear at the $\Gamma$ point, for which the chirality connected with nonzero PAM of $\pm 1$ under the $C_4$ rotation. Importantly, for chiral phonons at $\Gamma$, they can directly couple with optical transitions, either by resonant excitation or by Raman scattering, rather than the intervalley transition required before.
We perform a rough estimation for the effect in a realistic material, the 2D ferromagnetic MnAs monolayer, and find that the effect is typically weak due to the usually weak spin-lattice coupling.
Nevertheless, we show that by extending to 3D systems, chiral phonons with $C_4$ symmetry can appear on the high-symmetry path. In this case,  $\mathcal{T}$ breaking is not needed, but the lattice has to be chiral, as demonstrated by our first-principles calculation on $\alpha$-cristobalite.
Our work extends the concept of chiral phonons beyond the hexagonal crystal system, which offers platforms for studying chiral phonons with novel properties and potential applications.

\section{General analysis}

\begin{figure}
	\includegraphics[scale=0.4]{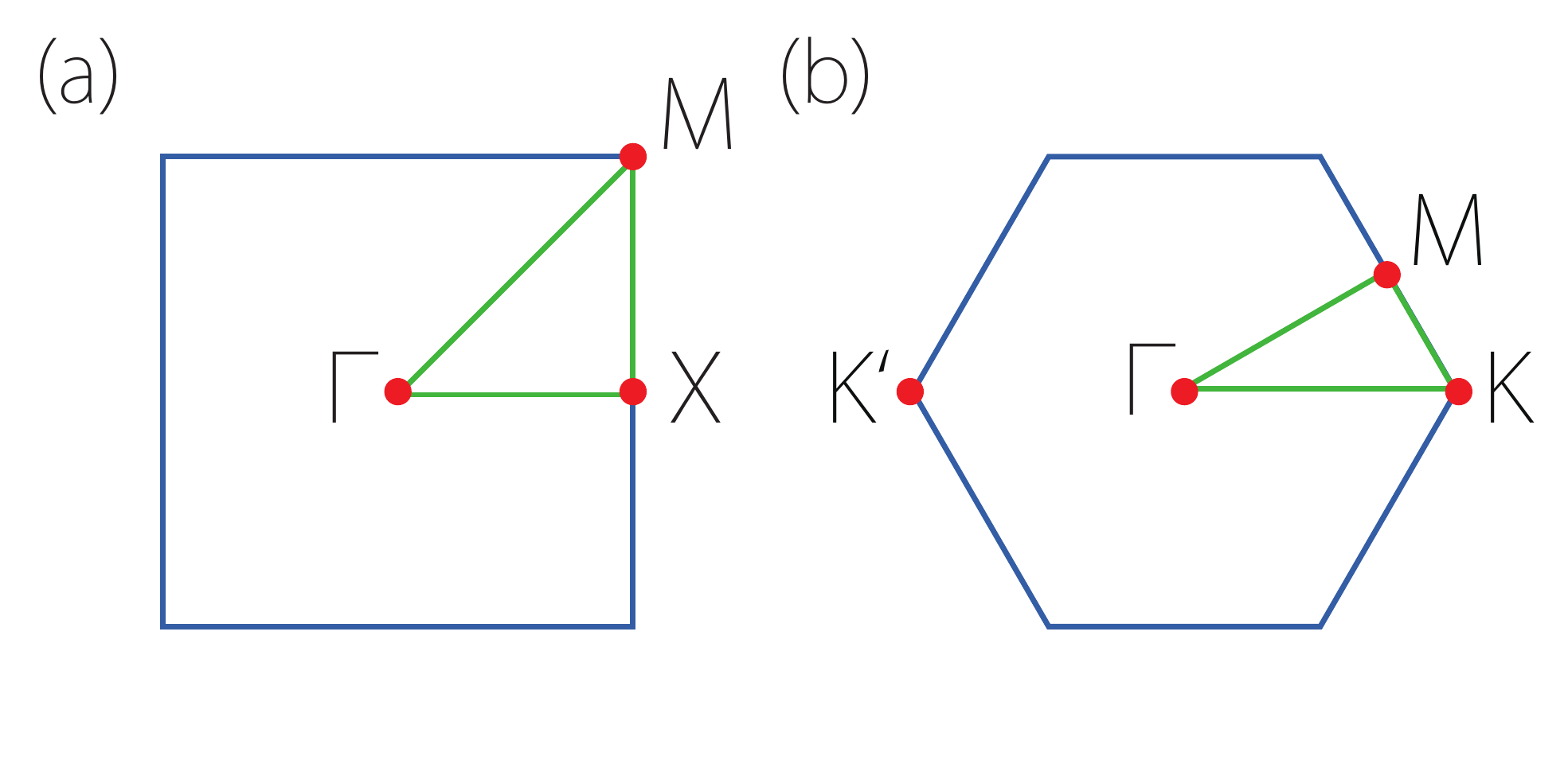}
	
	\caption{\label{fig:Figure1} Brillouin zones for (a) a square lattice and (b) a hexagonal lattice. }
\end{figure}

Let's first give a general consideration for 2D square lattices with $C_4$ symmetry. As discussed, our interest is on the possible chiral phonons at the high-symmetry points in BZ which respect the $C_4$ symmetry. These points are marked in Fig.~\ref{fig:Figure1}(a), which include the $\Gamma$ point and the $M$ $(\pi,\pi)$ point. The points $X$ and $Y$ are also high-symmetry points, but they generally retain only the $C_2$ symmetry. A crucial observation is that \emph{all} the high-symmetry points of the square lattice are TRIM points, i.e., $\mathcal{T}$ is a symmetry at these points.
It then follows that phonon modes at these points cannot have a net chirality. This is because the $\mathcal{T}$ operation flips the chirality of a phonon mode. Hence, a left-handed phonon mode must have a right-handed time reversal partner, and they are degenerate at the same energy, annihilating the net chirality (the degenerate pair can always be decomposed into nonchiral linear modes).

In comparison, for hexagonal class lattices studied before, their BZ is a hexagon (see Fig.~\ref{fig:Figure1}(b)). There are high-symmetry points, namely the $K$ and $K'$ points, which are not TRIM points. Therefore, we are allowed to have chiral phonon modes at $K$ and $K'$. Under $\mathcal{T}$, $K$ goes to $K'$ and vice versa. As a result, a left-handed mode at $K$ has its right-handed time reversal partner at $K'$. The separation in momentum space ensures their well defined chirality.

From this analysis, we see that a necessary condition to have chiral phonons in a square lattice is to break the $\mathcal{T}$ symmetry. Then, chiral phonons may emerge at the $\Gamma$ and $M$ points of the square BZ. Chiral phonons at $\Gamma$ are of particular interest, because they can be directly coupled with light, without worrying about the crystal momentum mismatch. In the following, we shall explore these ideas, first in a simple model and then in a realistic material.

\begin{figure}
	\includegraphics[scale=0.4]{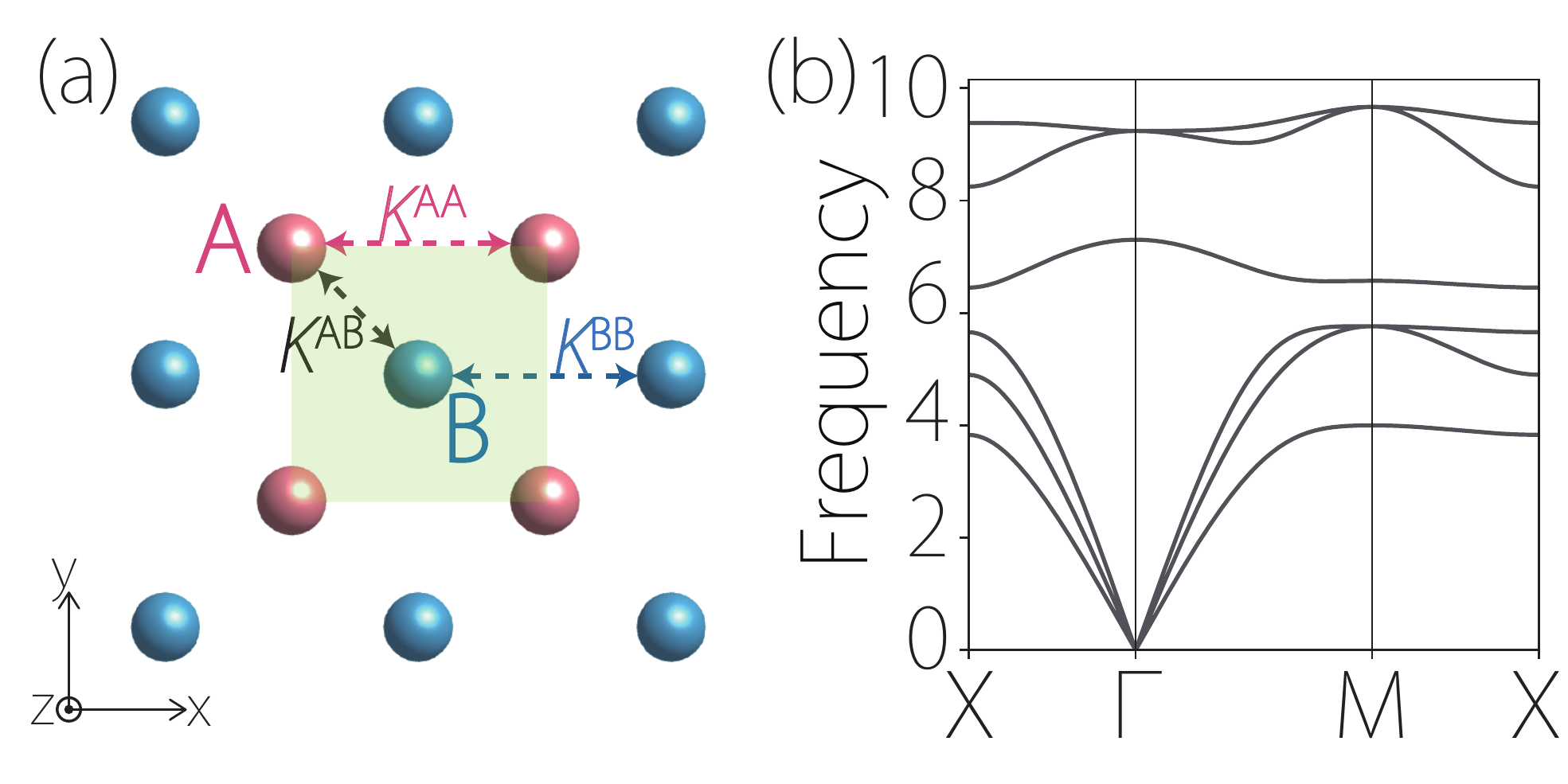}
	
	\caption{\label{fig:Figure2} (a) Schematic illustration of the  square lattice model. There are two basis sites A and B in a unit cell (marked by the shaded square). The couplings between AB, AA and BB sites are considered in the model.
		(b) Phonon dispersion of the model. In the calculation, we set the force constants
		$\{k_L^\beta, k_{Ti}^\beta,k_{To}^\beta\}$ for a bond $\beta=$ AB, AA and AB as  AB: $\{20,12,10\}$, AA: $\{8, 2, 1\}$, and BB: $ \{6, 1, 0.4\}$, respectively. }
\end{figure}

\section{2D square-lattice model\label{sec:2d_model}}

Let us consider the simple 2D square lattice model illustrated in Fig.~\ref{fig:Figure2}(a). Since we are interested in the chiral phonon modes in the optical branches, we need at least two atomic sites in a unit cell: A is at the corner of a square and B is at the center. The structure clearly preserves the $C_4$ symmetry.

The harmonic oscillations of the square lattice is described by the standard Hamiltonian:
\begin{equation}
	H_{0}=\frac{1}{2}p^{T}p+\frac{1}{2}u^{T}Ku\label{eq:H0}
\end{equation}
where $u$ is a column vector of displacements from lattice equilibrium positions for each atom, multiplied with the square root of atomic mass $m_{A/B}$ for A/B site; $p$ is the conjugate momentum vector, and $K$ is the force constant matrix.
For simplicity, in the model, we keep the couplings up to the second neighbors. These include the coupling through the AB, AA, and BB bonds, as indicated in Fig.~\ref{fig:Figure2}(a). Each bond $\beta$ is characterized by a longitudinal force constant $k_L^\beta$ and two transverse force constants $k_{Ti}^\beta$ (in-plane) and $k_{To}^\beta$ (out-of-plane). Since our target feature is dictated by symmetry, including more neighbors or variation in model parameters will not change the qualitative result.

The phonon modes $u_{k,\sigma}$  and the spectrum $\omega_{k,\sigma}$ are solved from the eigenvalue problem
\begin{equation}
	D(k\text{)}u_{k,\sigma}=\omega_{k,\sigma}^{2}u_{k,\sigma},\label{eq:D-frequency}
\end{equation}
where the dynamic matrix $D$ is the spatial Fourier transform of $K$, and the index $\sigma$ labels the phonon branches.

A representative spectrum of the model is shown in Fig.~\ref{fig:Figure2}(b). There are totally six phonon branches: three acoustic branches and three optical branches, in accordance with the two sites in a unit cell. Let's focus on the modes at the $\Gamma$ point. The three acoustic modes are degenerate at zero energy, as expected. As for the three optical modes, one is non-degenerate (the 4th branch), and the other two (the 5th and the 6th branches) form a degenerate pair.

To check the possible chirality of these phonons, we evaluate their circulation polarization~\cite{zhang2015chiral}. For a 2D system, we are interested in the circulation polarization along the out-of-plane ($z$) direction, which is given by
\begin{equation}
  s_{k,\sigma}=u_{k,\sigma}^\dagger \hat{S}_z u_{k,\sigma},
\end{equation}
where
\begin{equation}
  \hat{S}_z=\sum_\alpha (|R_\alpha\rangle\langle R_\alpha|-|L_\alpha\rangle\langle L_\alpha|)
\end{equation}
is the circular polarization operator, the summation is over all the sites in a unit cell, $|R_\alpha\rangle$ ($|L_\alpha\rangle$) is the right (left) circularly polarized vibration basis at site $\alpha$. Hence, $s_{k,\sigma}>0$ $(<0)$ will indicate that the mode $u_{k,\sigma}$ is right (left) handed, and it was shown that the value $\hbar s_{k,\sigma}$ gives the angular momentum of the phonon along $z$.

Straightforward evaluation confirms that the optical modes at $\Gamma$, including the non-degenerate mode and the doubly degenerate pair, have a vanishing chirality. This is consistent with our general analysis. The degenerate pair can be decomposed into a right-handed phonon and its left-handed time reversal partner, so they together have a zero net chirality. The similar discussion applies also to the other high-symmetry points.

Next, we break the $\mathcal{T}$ symmetry in this model and show that chiral phonons with $C_4$ symmetry can appear. Following Ref.\cite{zhang2010topological,holz1972phonons,komiyama2021universal}, we add a $\mathcal{T}$-breaking perturbation to the model, such that the Hamiltonian takes the form of
\begin{equation}
  H=\frac{1}{2}(p-\tilde{A}u)^{T}(p-\tilde{A}u)+\frac{1}{2}u^{T}Ku,\label{eq:HwithB LFZ}
\end{equation}
where $\tilde{A}$ is an anti-symmetric real matrix. Clearly, the term $p \tilde{A} u$ breaks the $\mathcal{T}$ symmetry. Physically, such a perturbation may result from different origins, such as the Lorentz force on charged ions~\cite{holz1972phonons}, Raman-type spin-phonon interaction~\cite{zhang2010topological}, or Coriolis force~\cite{kariyado2015manipulation,wang2015coriolis}. $\tilde{A}$ is block diagonal in the site indices $\alpha$, with each $3\times 3$ block $\Lambda_\alpha$ corresponding to the site $\alpha$ being given by
\begin{equation}\label{Lambda}
  \Lambda_i=\left(
              \begin{array}{ccc}
                0 & \lambda_\alpha & 0 \\
                -\lambda_\alpha & 0 & 0 \\
                0 & 0 & 1 \\
              \end{array}
            \right),
\end{equation}
with some constants $\lambda_\alpha$ signaling the strength of $\mathcal{T}$ breaking.
Intuitively, this resembles the effect of a Lorentz force acting on moving ions from an out-of-plane magnetic field, where $\lambda_\alpha$ would be proportional to the field strength.

\begin{figure}
	\includegraphics[scale=0.4]{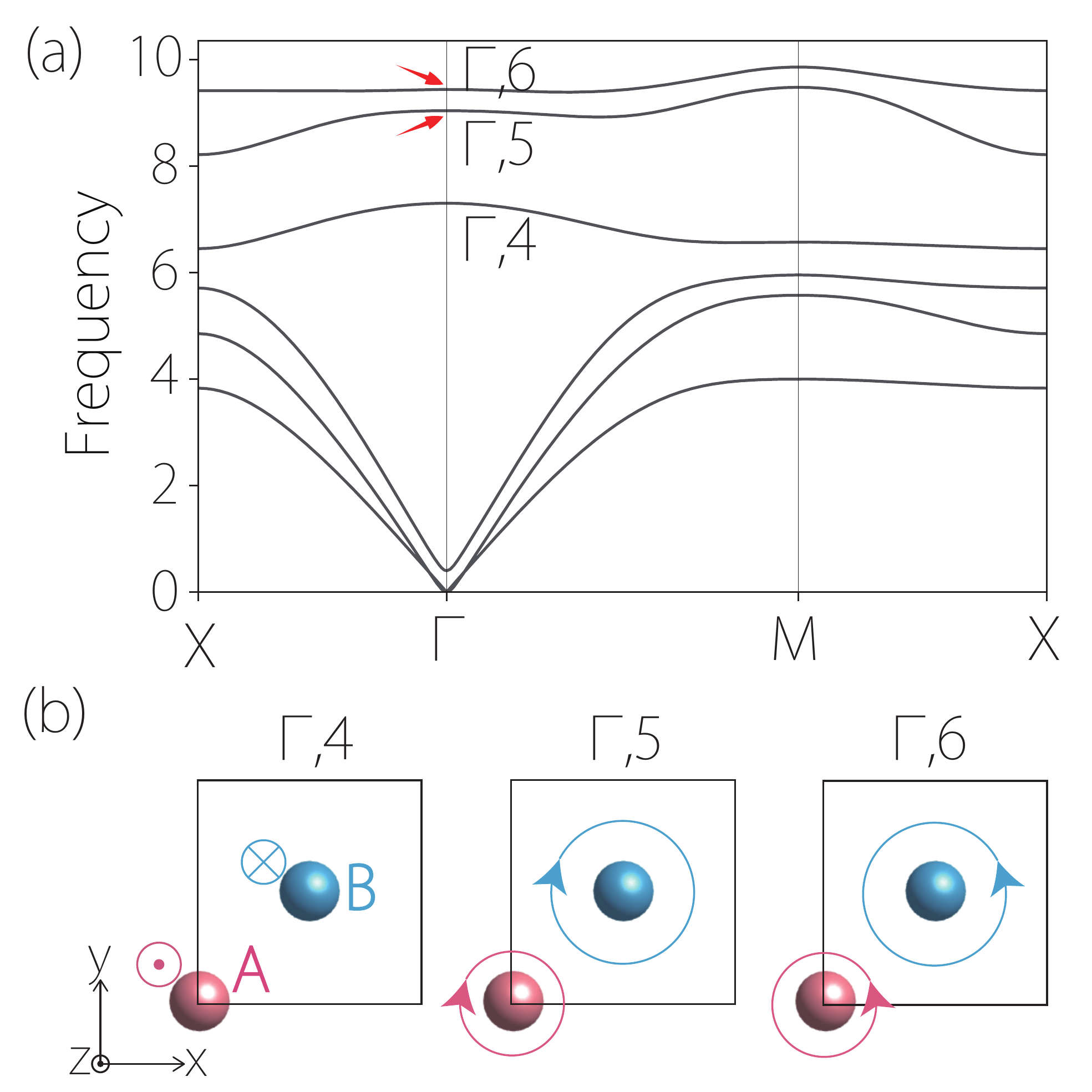}
	\caption{\label{fig:Figure3}(a) Phonon dispersion of the square lattice model with $\mathcal{T}$-breaking perturbation. The arrows mark the two $C_4$ chiral optical phonon modes at $\Gamma$. (b) Vibration patterns for the three optical phonon modes $u_{\Gamma,4}$, $u_{\Gamma,5}$ and $u_{\Gamma,6}$. }
\end{figure}

After adding the perturbation, the phonon spectrum is changed to that in Fig.~\ref{fig:Figure3}(a). Compared to Fig.~\ref{fig:Figure2}(b), one finds that several degeneracies in the spectrum are lifted by the perturbation. At the $\Gamma$ point, one acoustic mode splits from the other two and acquires a finite gap, which was noted in a recent work~\cite{komiyama2021universal}. More importantly, for the optical modes, the original doubly degenerate pair gets separated. By evaluating their circular polarization, we find that
{the mode $u_{\Gamma,5}$ of the 5th branch is left handed, whereas $u_{\Gamma,6}$ is right handed.}
Their chirality would flip when $\tilde{A}$ changes sign, as it should be.

The phonon chirality can be directly visualized from the vibration pattern, as plotted in Fig.~\ref{fig:Figure3}(b) for the three optical modes at $\Gamma$.
{One can see that the mode $u_{\Gamma,4}$ is non-chiral as vibrations of the two sites are linear and along the $z$ direction. For $u_{\Gamma,5}$, the two sites perform left handed circular rotation around their equilibrium positions, consistent with the left handed chirality. Meanwhile, $u_{\Gamma,6}$ has similar circular vibration pattern as $u_{\Gamma,5}$, except that orientation is right handed.}
The two are connected by the $\mathcal{T}$ operation, hence they are degenerate when $\mathcal{T}$ is preserved. These results confirm that by breaking the time reversal symmetry, we can indeed have chiral phonons in a square lattice with $C_4$ symmetry.

As mentioned, due to $C_4$, the modes at $\Gamma$ have a well defined PAM. The PAM is determined by the $C_4$ eigenvalue of the phonon mode. Explicitly, we have
\begin{equation}\label{PAM}
  \mathcal{R}_z(\pi/2) u_{k,\sigma}=e^{-i(\pi/2)\ell_{k,\sigma}} u_{k,\sigma},
\end{equation}
where $\mathcal{R}_z(\pi/2)$ is the $C_4$ operator acting on the phonon wave function, $k\in\{\Gamma, M\}$ here is restricted to the $C_4$ invariant points, and the PAM $\ell \in\{0,\pm 1, 2\}$.
{In Table~\ref{tab:Mag_Model},}
we show the PAM values for the three optical modes at $\Gamma$.
One observes that the right-handed (left-handed) chiral phonon has PAM of $+1$ ($-1$). For this simple lattice, the PAM is directly connected to the chirality.

To better understand the PAM, we note that as discussed in Ref.~\cite{zhang2015chiral}, the phase factor on the right hand side of Eq.~(\ref{PAM}) has two contributions: an intracell contribution from the vibration at a site $i$ and an intercell contribution from the Bloch phase factor $e^{i\bm k\cdot \bm{r}}$ when site $\alpha$ is moved to neighboring cell under rotation. Regarding their contributions to the total PAM $\ell$, the former is termed as the spin PAM ($\ell_s$) and the latter is termed as the orbital PAM ($\ell_o$). And the relation $\ell=\ell_s^\alpha+\ell_o^\alpha$ holds for each site $\alpha$. Now, at the $\Gamma$ point, the orbital PAM vanishes identically because $\bm k=0$. Hence, the PAM is completely contributed by the spin PAM, which is determined by the vibration pattern at any $C_4$ invariant site (A or B here). It then follows that PAM of $\pm 1$ corresponds to left/right handed phonon in our square lattice model.

\begin{table}[tb]
	\caption{\label{tab:Mag_Model}%
		Results for optical phonon modes at $\Gamma$ and $M$ points in Fig.~\ref{fig:Figure3}(a). Here, the columns with ``$C_{4}$" and ``$\ell_{ph}$" give the $C_4$ eigenvalues and phonon PAM, respectively. R/L indicates the right/left handed chirality.  }
	\begin{ruledtabular}
		\begin{tabular}{ccccccc}
			&
			\multicolumn{3}{c}{$\Gamma$}  &
			\multicolumn{3}{c}{$M$} \tabularnewline
			\cline{2-4}	\cline{5-7}
			&$C_{4}$& $\ell_\text{ph}$ & chirality & $C_{4}$ & $\ell_{ph}$ & chirality \\
		   	\hline
		    $\Gamma$,6 & $-i$ & 1 & R &  $-i$ & 1 & R \\
		    $\Gamma$,5 & $i$ & $-1$ & L & $i$ & $-1$ & L \\
		    $\Gamma$,4 & 1 & 0 & - & 1 & 0 & -\\
		\end{tabular}
	\end{ruledtabular}
\end{table}

We have also investigated chiral phonons at the other $C_4$-symmetric point $M$.
{The results are shown in Table~\ref{tab:Mag_Model}.}

Chiral phonons at $\Gamma$ point have the advantage that they can directly couple with light. Because the wavelength of light in the infrared to visible range is much larger than the crystal lattice scale, its momentum can only match the phonon modes at very small $k$, i.e., around the $\Gamma$ point. In previous experiments, to probe chiral phonons at $K$ and $K'$ points of the hexagonal 2D transition metal dichalcogendies materials, one has to invoke an intervalley scattering process, where the momentum mismatch is compensated by an electron. Now, if we have chiral phonons at $\Gamma$ point, they may be resonantly excited by an infrared circularly polarized light, which follows the selection rule
\begin{equation}\label{select}
  \ell_\text{ph}=m,
\end{equation}
where $\ell_\text{ph}$ is the PAM of the target chiral phonon mode, and $m=\pm 1$ for right/left circularly polarized light. In addition, for chiral modes that are Raman active (not in this simple model), they can also be probed in the Raman spectrum.

\section{Calculation for a real material}

We have shown that it is possible to have chiral phonons in a square lattice with $C_4$ symmetry. A necessary condition is that the $\mathcal{T}$ symmetry must be broken. In solid materials, $\mathcal{T}$ may be broken by external magnetic field or by internal magnetic ordering. Here, we perform a calculation on a realistic material, the monolayer MnAs. The details of the calculation are presented in Appendix ~\ref{sec:Appendix_A}.

\begin{figure}
	\includegraphics[scale=0.4]{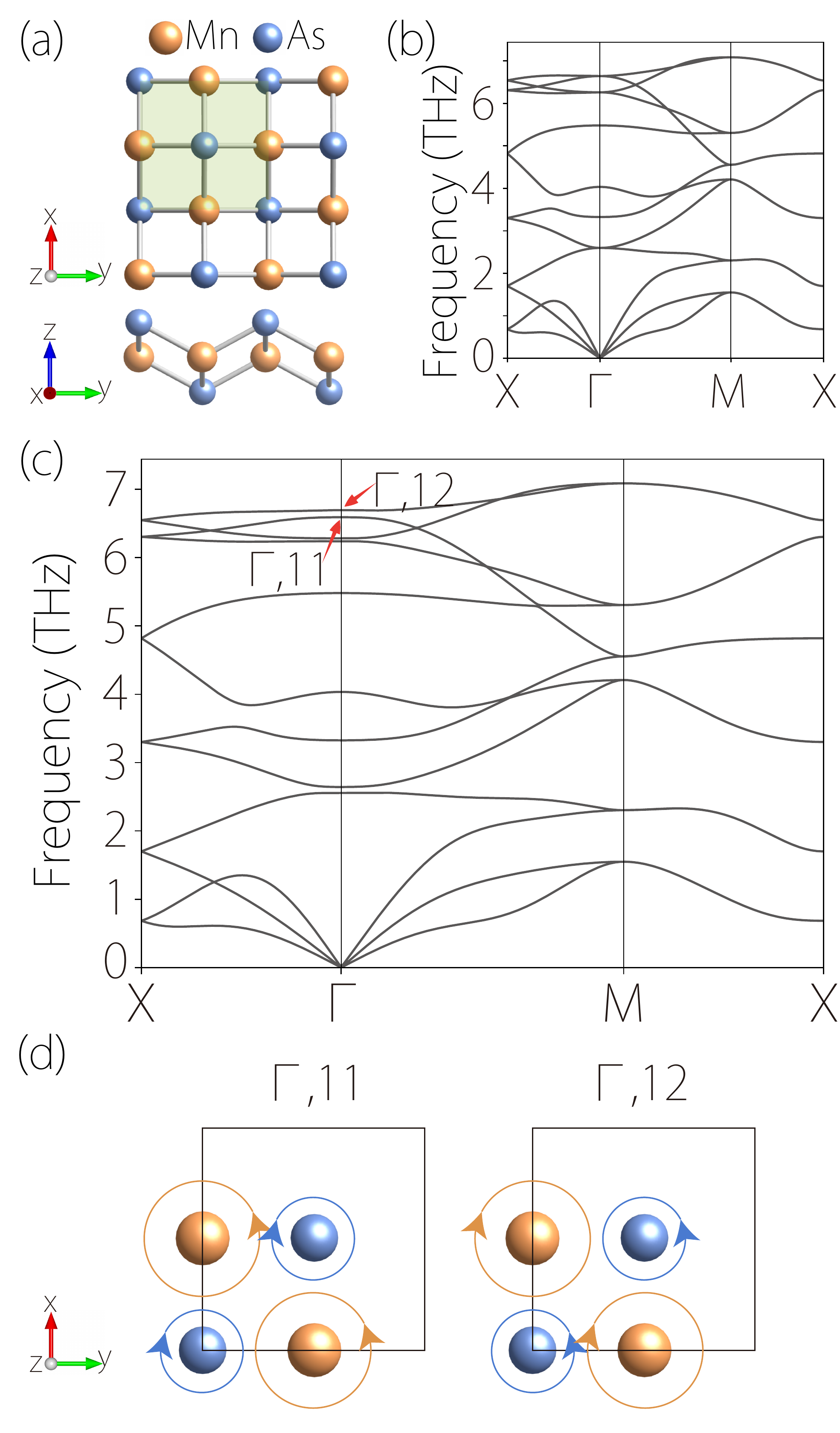}
	
	\caption{\label{fig:Figure4}
		(a) Top and side views of MnAs monolayer.
		(b) Calculated phonon dispersion of MnAs monolayer without $\mathcal{T}$ breaking.
		(c) The corresponding dispersion by taking a $\mathcal{T}$-breaking perturbation with $B=10^5$ T. One notes the splitting of the original
double degeneracy at $\Gamma$ in (b). The resulting split modes are chiral. The arrows indicate two such split modes (corresponding to 11th and 12th branches), and their vibration patterns are illustrated in (d).}
\end{figure}

The structure of monolayer MnAs is shown in Fig.~\ref{fig:Figure4}(a) It adopts the FeSe-type structure with space group $P4/nmm$ (No.~129) and point group $D_{4h}$. The Mn atoms form a horizontal plane, and the As atoms are located on the two sides of this plane. Each Mn is sitting inside a tetrahedron formed by four neighboring As atoms. A unit cell contains two Mn and two As atoms. As proposed by Wang \emph{et al.}~\cite{wang2019mnx}, monolayer MnAs is intrinsically a half metal with out-of-plane magnetization. Our first-principles calculations (see Appendix \ref{sec:Appendix_A} for details) confirm this ferromagnetic ground state. The local moment is mainly from the Mn $3d$ orbitals and is about $4 \mu_B$ per Mn site. Notably, the out-of-plane magnetization preserves the $C_4$ symmetry of the system and breaks the $\mathcal{T}$ symmetry.

In Fig.~\ref{fig:Figure4}(b), we plot the calculated phonon spectrum of the system without considering the $\mathcal{T}$-breaking effects. There are 12 phonon branches, corresponding to the four atoms in a unit cell.  Focusing on the optical phonon modes at the $\Gamma$ point, we see that these modes are either non-degenerate or doubly degenerate. Each doubly degenerate pair corresponds to a two-dimensional irreducible representation ($E_g$ or $E_u$) of the $D_{4h}$ group. As expected from our general analysis, it can be regarded as consisting of a right-handed mode and a left-handed mode, which are connected by the time reversal operation, so the net chirality vanishes. The non-degenerate modes correspond to out-of-plane vibrations, which are also non-chiral. These points are explicitly verified by the calculation of the phonon circular polarization $s_\text{ph}$ for these modes, as shown in Table~\ref{tab:MnAs}.

\begin{table}[b]
	\caption{\label{tab:MnAs} Results for optical phonon modes at the $\Gamma$ point in MnAs monolayer. The frequency $\omega$ is in unit of THz, IRR shows the irreducible representation of the modes,  and $s_\text{ph}$ is the phonon circular polarization. R/L indicates the right/left handed chirality. The left panel is for the results without $\mathcal{T}$ breaking (Fig.~\ref{fig:Figure4}(b)), where the modes do not have a net chirality. The right panel is for the case with $\mathcal{T}$ breaking perturbation from a $B$ field of $10^5$ T (Fig.~\ref{fig:Figure4}(c)).}
	\begin{ruledtabular}
		\begin{tabular}{cccccc}
			\multicolumn{3}{c}{Without $\mathcal{T}$ breaking} &
			\multicolumn{3}{c}{$B=10^{5}$ T}\tabularnewline
			\cline{1-3} \cline{4-6}
			
			$\omega$ & IRR & $\ell_\text{ph}$ & $\omega$   & $\ell_\text{ph}$ & $s_\text{ph}$\tabularnewline
			\cline{1-3} \cline{4-6}
			
			\multirow{2}{*}{6.644} & \multirow{2}{*}{$E_{g}$} &  \multirow{2}{*}{(1,$-1$)} &
			6.696 &  1 & L\tabularnewline
			
			 &  &  &  6.593 & $-1$ & R\tabularnewline
						
			\multirow{2}{*}{6.261} & \multirow{2}{*}{$E_{u}$} &  \multirow{2}{*}{(1,$-1$)} &
			6.283 &  $-1$ & L\tabularnewline
			
			&  &  &    6.240 & 1 & R\tabularnewline
					
			5.482 & $A_{2u}$  & 0   & 5.482  & 0 & -\tabularnewline
			
			4.034 & $A_{1g}$  & 0   & 4.034  & 0 & -\tabularnewline
			
			3.325 & $B_{1g}$  & 2   & 3.325 &  2 & -\tabularnewline
			
			\multirow{2}{*}{2.598} & \multirow{2}{*}{$E_{g}$} &  \multirow{2}{*}{(1,$-1$)} &
			2.642 &  1 & R\tabularnewline
			
			&  &  &   2.554 &  $-1$ & L\tabularnewline
		\end{tabular}
		
	\end{ruledtabular}
\end{table}

Next, we try to include $\mathcal{T}$ breaking effects on phonons. We note that currently, there is no established method to include such effects in first-principles calculations. A recent work by Sun \emph{et al.}~\cite{sun2021phonon} proposed an attempt to include
magnetic field effects on phonons, based on adding the spin-phonon interaction term as in model~(\ref{eq:HwithB LFZ}). The block $\Lambda_\alpha$ for site $\alpha$ in the $\tilde{A}$ matrix is expressed by
\begin{equation}
  \bm\Lambda_\alpha=\frac{e}{4m_{\alpha}}({\bm Z}_{\alpha}^{T}\times \bm B+\bm B\times \bm Z_{\alpha}),\label{eq:A-born-charge}
\end{equation}
where $m_\alpha$ is the mass of the ion at site $\alpha$, $\bm Z_\alpha$ is its Born effective charge dyadic, $\bm B$ is the magnetic field, and here the matrix $\Lambda_\alpha$ is also expressed in the dyadic form: $\bm\Lambda=\sum_{ij}\Lambda_{ij}\bm e_i \bm e_j$ with $\bm e_i$ the Cartesian basis vectors. For the special case when the field is along $z$ and the Born effective charge tensor is given by the simple product of some charge $q_\alpha$ and the identity matrix, $\Lambda_\alpha$ would reduce to the form in Eq.~(\ref{Lambda}) with $\lambda_\alpha=-q_\alpha B/(2m_\alpha)$.

We follow this approach and perform the calculation for monolayer MnAs. We first consider the effect from an external $B$ field. Figure~\ref{fig:Figure4}(c) shows the obtained phonon spectrum for a $B$ field of $10^5$ T along the $+z$ direction. Consistent with our expectation, one observes that by breaking the $\mathcal{T}$ symmetry, the double degeneracies in Fig.~\ref{fig:Figure4}(c) for the optical modes at $\Gamma$ are lifted. Then, each split mode from the original degeneracy carries a net chirality and well defined PAM. These values are presented in Table~\ref{tab:MnAs}. For example, the top two modes at $\Gamma$ evolve from the original degeneracy at {$\omega=6.644$ THz.}
They are left and right handed, respectively. The in-plane vibration patterns of these two modes are illustrated in Fig.~\ref{fig:Figure4}(d).
Similar analysis can be done also for the phonon modes at $M$, and the results are consistent with our general consideration in Sec.~\ref{sec:2d_model}.

Clearly, the splitting between the chiral modes scales linearly with the field strength. We note that from our calculation, sizable splitting of the degeneracy only occurs at very large field strength. For example, in Fig.~\ref{fig:Figure4}(c), the splitting is on the order of {0.1 THz} at $B$ field of $10^5$ T. The similar behavior was also observed in Ref.~\cite{sun2021phonon}. Evidently, such huge magnetic field cannot be achieved under current lab condition. Since monolayer MnAs is a ferromagnetic material, the internal magnetization  breaks $\mathcal{T}$ and should also produce a splitting. However, there is so far no developed approach to capture this effect in first-principles calculations. Here, we may do a very rough estimation by attributing the spin splitting in the material (splitting between two spin channels) to an ``internal" $B$ field and taking this field in Eq.~(\ref{eq:A-born-charge}). From the calculated band structure (see Appendix ~\ref{sec:Appendix_B}), the spin splitting is found to be $\Delta\sim 2 $ eV, hence $B\sim \Delta/\mu_\text{Mn}\sim 10^4$ T. As a result, the splitting in phonon spectrum is at least one order of magnitude smaller than that in Fig.~\ref{fig:Figure4}(c).

\section{$C_4$ chiral phonons in 3D systems}

From the above discussion, we see that chiral phonons with $C_4$ symmetry can exist in 2D square lattices when the $\mathcal{T}$ symmetry is broken.
However, the estimation with monolayer MnAs indicates that effects of $\mathcal{T}$ breaking on phonons could be rather weak, such that the splitting between modes with opposite chirality could be difficult to detect  under current lab condition.

\begin{figure}
	\includegraphics[scale=0.4]{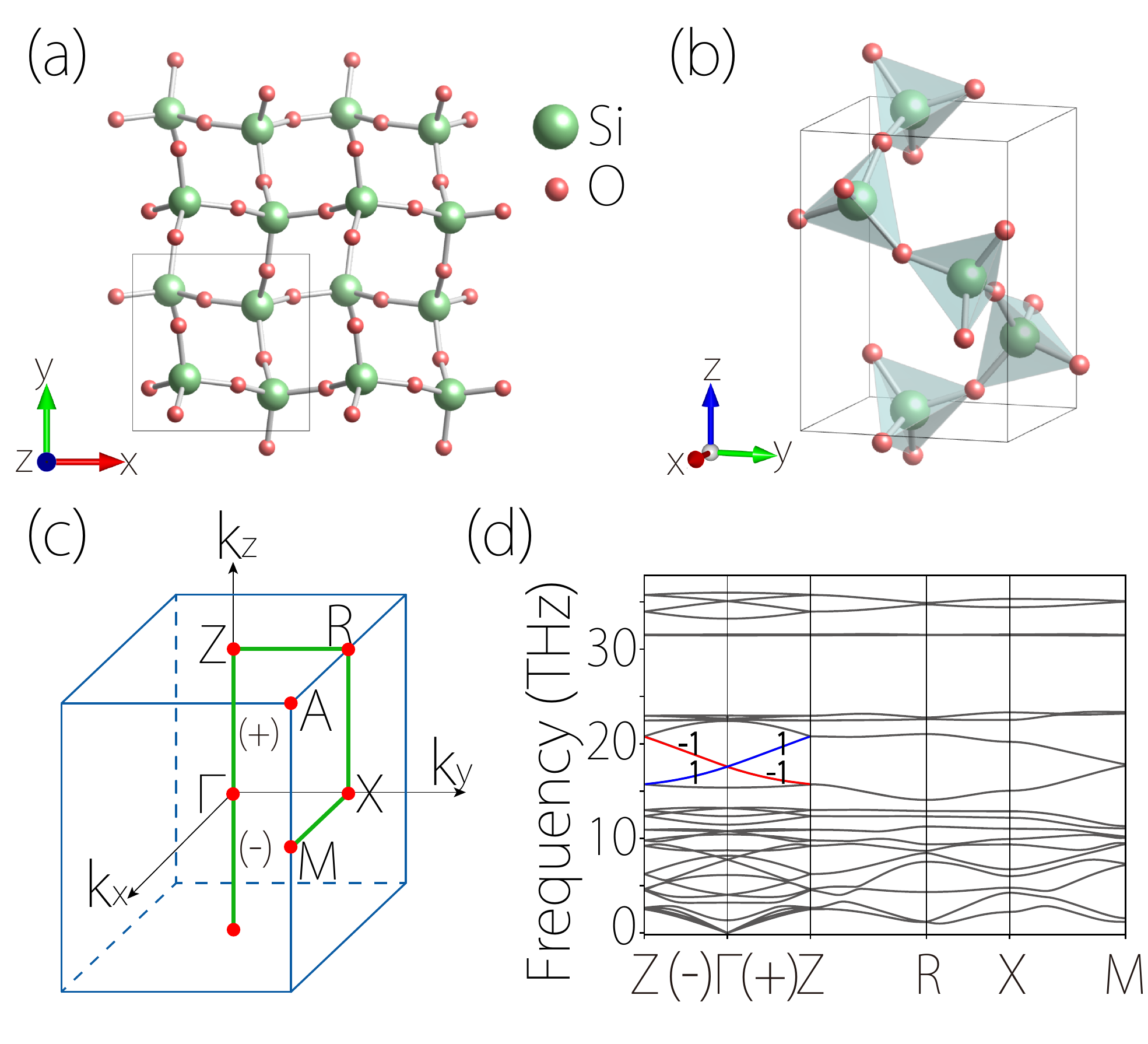}
	\caption{\label{fig:Figure5}
		(a) Top view of the $\alpha$-cristobalite lattice, which has tetragonal symmetry.
		(b) A chain of SiO$_{4}$ tetrahedra along the $z$ direction, showing the pattern of a left handed spiral.
		(c) Brillouin Zone. Here, the high-symmetry path $\Gamma$-$Z$ with $k_z>0$ and $k_z<0$ are labeled with $(+)$ and $(-)$ signs, since they are not equivalent regarding the properties of phonon modes.
		(d) Calculated phonon dispersion for $\alpha$-cristobalite.
		Here, we focus on the two phonon branches within the range of 17 - 22 THz. The red/blue color indicates the mode is right/left handed. The values $\pm 1$ indicate the PAM.}
\end{figure}

The problem may be circumvented when extending the discussion to 3D systems. For a 3D lattice with $C_4$ symmetry, i.e., a tetragonal lattice, the $C_4$ symmetry is preserved on the whole path $\Gamma$-$Z$ of the BZ, not just the high-symmetry points, as shown in Fig.~\ref{fig:Figure5}(c).
Note that $\mathcal{T}$ is not a symmetry for a generic point on the path, so there is no degeneracy caused by $\mathcal{T}$ for chiral modes at such a point and hence no need to break $\mathcal{T}$. In other words, chiral phonons can appear on $C_4$ invariant paths for 3D lattices that preserve the time reversal symmetry. Nevertheless, for such cases, some crystal symmetries must be broken. In particular, a necessary condition is that the system cannot preserve any mirror symmetry. This can be easily understood by noting that any vertical mirror (like $M_x$) or the combined symmetry $\mathcal{T}M_z$ would enforce degeneracy between left and right handed phonons on the $C_4$ invariant paths. In retrospect, this condition should also be met in 2D systems, and indeed, in our previous examples, these symmetries are automatically broken by the $B$ field or magnetization.

To demonstrate our idea, we perform the calculation for a 3D material
{$\alpha$-cristobalite} that satisfies the above condition. $\alpha$-cristobalite is a well known polymorph of silica (SiO$_2$)~\cite{bates1972raman}.  
Note that a crystal that lacks any mirror symmetry belong to the chiral space group. $\alpha$-cristobalite crystallizes in a pair of enantiomorphic tetragonal chiral space groups $P4_{1}2_{1}2$ (No.~92) and $P4_{3}2_{1}2$ (No.~96).
Figure~\ref{fig:Figure5}(a) and (b) show the structure of $\alpha$-cristobalite in the space group No.~92. Here, each Si atom stays in a tetrahedron of four nearby O atoms and has a tetrahedral coordination, and two nearby tetrahedra are connected at a corner O atom. From Fig.~\ref{fig:Figure5}(b), one can see that
{the SiO$_{4}$ tetrahedra form a left handed spiral chains  along $z$. }
As for the enantiomorphic structure in space group No.~96, the handedness of the spiral would be the opposite. Importantly, the structure preserves a fourfold screw rotation along $z$, which allows us to discuss $C_4$ chiral phonons with well defined PAM.

The calculated phonon dispersion of $\alpha$-cristobalite is plotted in Fig.~\ref{fig:Figure5}(d) (see Appendix ~\ref{sec:Appendix_A} for computational details). There are 36 phonon branches, corresponding to the four formula units in a primitive cell. As discussed, our target here is on the phonon modes on the $\Gamma$-$Z$ path.
{The little group on this path is $C_{4}$.}
The modes on this paths are generally non-degenerate, except for some accidental crossing points. We have checked that they indeed have nonzero phonon circular polarization (along $z$) and well defined PAM. For example, let's focus on the two branches from about 17 to 22 THz, since they are well separated from other branches. In Fig.~\ref{fig:Figure5}(d), we mark their chirality by colors: red for right-handed modes and blue for left-handed modes. The PAM values for these modes are also labeled in the figure. One observes that as expected, the $C_4$ phonons on this paths are chiral and have PAM of $\pm 1$. Moreover, as noted in Ref.~\cite{chen2021chiral}, for a chiral crystal, the phonon chirality is tied with its propagation direction. Here, the blue colored branch is propagating in the $+z$ direction, whereas the red colored branch goes in the opposite direction.

\section{Discussion and conclusion}

In this work, we have extended chiral phonons to square/tetragonal crystal systems with fourfold rotational symmetry. In 2D, to have $C_4$ phonons with net chirality, a necessary condition is to break the time reversal symmetry. Our estimation shows that the phonon splitting due to symmetry breaking effects from applied magnetic field or magnetic ordering could be rather weak for realistic materials under currently achievable lab conditions. For example, recent experiments on hexagonal magnets did not resolve the phonon splitting due to magnetic ordering~\cite{yin2021chiral,du2019lattice}.
Nevertheless, we have to stress that the estimation is very crude. So far, we do not have a good microscopic theory to account for the $\mathcal{T}$ breaking effects on phonons, and to capture such effects in first-principles calculations is an important open problem to be explored in future research. We hope our current work provides an additional stimulus for the development.

We have shown that in 3D, $C_4$ chiral phonons can appear on the high-symmetry path of a chiral tetragonal crystal, without the need to break $\mathcal{T}$. For these chiral phonons, the optical selection rule in (\ref{select}) still holds. For example, consider the chiral phonons in Fig.~\ref{fig:Figure5}(d) for $\alpha$-cristobalite and an incident light along $+z$. If the light is peaked around 18 THz, it will primarily interact with the two colored optical phonon branches. Then, the light with left (right) circular polarization can only resonantly excite left (right) handed phonon branch. Since the phonon chirality is tied to the propagation direction, this selectivity can be detected experimentally by the different heat flow direction.

Finally, although our discussion is mainly on phonons in solid materials, the analysis from symmetry perspective is general and also applies to artificial systems such as acoustic crystals and mechanical networks. Some effects may be more pronounced and more easily realized in artificial systems. For example, the $\mathcal{T}$ breaking may be achieved in artificial systems by other means, such as Coriolis force or optomechanical coupling~\cite{liu2020topological}.

In conclusion, we have explored chiral phonons beyond the hexagonal lattice systems. We show that $C_4$ chiral phonons can in principle exist. We clarify the required symmetry conditions for both 2D and 3D systems. For 2D, $C_4$ chiral phonons require broken $\mathcal{T}$, which could be stringent for real materials. For 3D, the condition is less stringent but requires a chiral tetragonal crystal structure. These phonons have the advantage that they can directly couple with light, which would facilitate the experimental study. Our work enriches the fundamental understanding of chiral phonons in a new crystal system and offers a foundation for further investigating their interesting physical properties.

\begin{acknowledgments}
		The authors thank D. L. Deng for helpful discussions. This work is supported by Singapore Ministry of Education AcRF Tier 2 (MOE2019-T2-1-001) and National Research Foundation (NRF-CRP22-2019-0004). We acknowledge computational support from the Texas Advanced Computing Center and the National Supercomputing Centre Singapore.
\end{acknowledgments}

\appendix

\section{First-principles Computation Method\label{sec:Appendix_A}}

We performed first-principles calculations to study the electronic and phononic properties of MnAs and $\alpha$-cristobalite. The calculations were done based on the density functional theory (DFT) as implemented in the Vienna ab initio simulation package (VASP)~\cite{kresse1993ab,kresse1999ultrasoft}. The exchange-correlation functional was modeled within the generalized gradient approximation (GGA) with the Perdew-Burke-Ernzerhof realization~\cite{perdew1996generalized}. The projector augmented wave method~\cite{blochl1994projector} was adopted. The plane-wave cutoff energy was set to 500 eV. The energy and the force convergence criteria were set to be $10^{-7}$ eV and $10^{-2}$ eV$/\text{\AA}$, respectively. The phonon spectra were obtained by using the density functional perturbation theory (DFPT) method and the PHONOPY code~\cite{togo2015first}.  
For MnAs, the GGA + U method~\cite{anisimov1991band} with $U_{\text{eff}}=4$ eV was applied for the $d$ orbitals of Mn atoms to describe the strong correlated interaction of $d$ electrons. The $3\times3\times1$ supercell and a size of $3\times3\times1$ $\Gamma$-centered k mesh~\cite{monkhorst1976special} in the BZ were used for the phononic calculation. 
For $\alpha$-cristobalite, to obtain the phonon spectra, the $2\times2\times2$ supercell was used with a $3\times3\times3$ $\Gamma$-centered k mesh~\cite{monkhorst1976special} in the BZ. The method of non-analytical term correction (NAC)~\cite{gonze1997dynamical} was applied to get the dynamical matrix for $\alpha$-cristobalite.

\section{Electronic Band Structure for MnAs Monolayer\label{sec:Appendix_B}}

\begin{figure}[b]
	\includegraphics[scale=0.46]{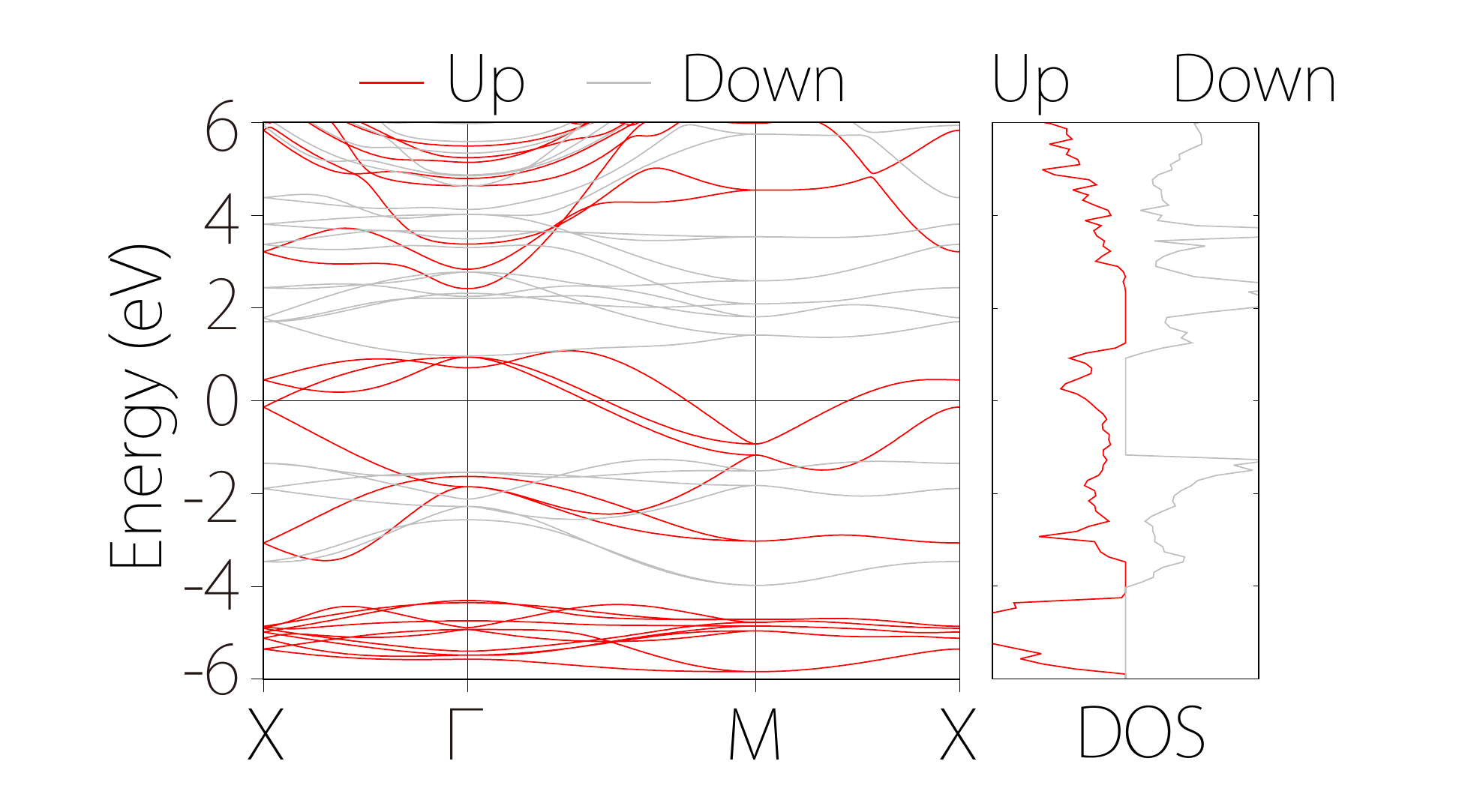}
	\caption{\label{fig:FigureA1}
		Electronic band structure and spin-resolved density of states for MnAs monolayer. }
\end{figure}

The electronic band structure for monolayer MnAs obtained from our DFT calculation is shown in Figure~\ref{fig:FigureA1}. From the spin-resolved density of states, the spin splitting in the material is found to be $\sim 2$ eV. 

%

\end{document}